\documentclass[amssymb,aps,twocolumn,showpacs,nofootinbib]{revtex4-1}
\usepackage[]{graphicx}
\usepackage[]{amsmath}
\usepackage{hyperref}
\usepackage{color}
\input{Qcircuit}

\def\ket#1{| #1 \rangle}

\def\eq#1{Eq.~\eqref{eq:#1}}

\def\eq#1{Eq.~\eqref{eq:#1}}

\begin{document}

\title{Reducing the quantum computing overhead with complex gate distillation}
\author{Guillaume Duclos-Cianci}
\affiliation{D\'epartement de Physique, Universit\'e de Sherbrooke, Qu\'ebec, Canada}
\author{David Poulin}
\affiliation{D\'epartement de Physique, Universit\'e de Sherbrooke, Qu\'ebec, Canada}

\date{\today}

\begin{abstract}
In leading fault-tolerant quantum computing schemes, accurate transformation are obtained by a two-stage process. In a first stage, a discrete, universal set of fault-tolerant operations is obtained by error-correcting noisy transformations and distilling resource states. In a second stage, arbitrary transformations are synthesized to desired accuracy by combining elements of this set into a circuit. Here, we present a scheme which merges these two stages into a single one, directly distilling complex transformations. We find that our scheme can reduce the total overhead to realize certain gates by up to a few orders of magnitude. In contrast to other schemes, this efficient gate synthesis does not require computationally intensive compilation algorithms, and a straightforward generalization of our scheme circumvents compilation and synthesis altogether.  
\end{abstract}


\maketitle


The accuracy threshold theorem \cite{Sho96a,AB96a,Kit97b,KLZ98b,Pre98a} states that if a physical device can realize one- and two-qubit operations to an accuracy of approximately 1\% \cite{Kni05a,RH07a}, then fault-tolerant techniques can be used to reliably quantum-compute with this device for arbitrary long times. This comes at the cost of consuming additional gates and qubits, but in principle this overhead grows `only' polynomially with the logarithm of the duration of the algorithm.  While there are today a few architectures with accuracies near or below threshold, e.g. \cite{DS13a,SNMB13a}, fault-tolerant quantum computing remains elusive, and a major bottleneck is the prohibitive fault-tolerance overhead. Part of the problem is that the devices' accuracies are too close to the threshold; they should ideally operate one or two orders of magnitude below threshold. But even in such ideal circumstances, the overhead would remain excessively high due to the cost of distillation and gate synthesis.

Because of their continuous nature, it is not possible to error-correct arbitrary quantum operations. Instead, fault-tolerant schemes realize a finite set of discrete, near-perfect universal operations. This universal set of fault-tolerant operations (USFTO) typically includes Clifford operations, since they are naturally fault-tolerant in may encoding schemes, e.g.  \cite{Sho96a,RH07a,B10a1}. Adding any non-Clifford operation to this set renders it universal. Magic state distillation and injection \cite{BK05b} is amongst the most efficient ways to generate these non-Clifford operations. 

State injection appends an ancillary register prepared in a magic state to the data register, performs a Clifford transformation on the joint registers, and measures a Pauli operator on the ancillary register. The resulting effect on the data register is a transformation $R(m)$ which depends on the measurement outcome $m$. A concrete example is detailed below, cf. Fig. 1 a).  Near-perfect magic states are obtained from noisy ones using {\em state distillation}, a process that uses only Clifford operations. Distilling a magic state to accuracy $\epsilon$ requires a number of noisy input states which  grows `only' polynomially with $\log (1/\epsilon)$, but even with our best distillation protocols this cost remains substantial \cite{MEK12a,DS12b,BH12b,J13c}.

Operations from this USFTO can be assembled to approximately {\em synthesize} any logical gate to accuracy $\delta$. The cost of synthesizing increases `only' polynomially with $\log(1/\delta)$ \cite{Sol00a,Kit97b}, but again for realistic applications, this cost is excessively large. Moreover, unlike  for the error-correction and distillation overheads, improving the physical devices is of no help; only software improvements can reduce the gate synthesis overhead. This is the problem of gate {\em compiling}. Naturally, compiling algorithms which use an exponential amount of classical computation  achieve shorter gate synthesis circuits \cite{F11a,PS13a,KMM12a}, although an efficient and optimal compiler has recently been discovered for circuits that make use of no ancillary qubits \cite{RS14a}. 

In this Letter, we present a scheme to distill a rich family of quantum transformations, which offers several advantages. 1) The total overhead of our scheme can be a few orders of magnitudes lower than what is achieved combining state-of-the art distillation and synthesis techniques. 2) This is achieved by an efficient compilation algorithm. 3) A generalization of our scheme can reduce the gate synthesis cost of any single-qubit gate to a constant.

To get a sense of the overhead associated to distillation and synthesis, suppose that the quantum algorithm we are executing requires $10^4$ logical qubits and has a circuit depth of $10^6$, roughly the size required to factor a 1024 bit integer \cite{MLFY10a} as used in common encryption schemes. This entails $10^{10}$ events where errors can occur, so each component in the circuit needs to be accurate to at least 10 digits to prevent imperfections from building up and scrambling the information. Assuming that Clifford operations can be executed perfectly (thus ignoring the error-correction cost as is usually done in such analysis) and using state-of-the-art compiling sequences \cite{KMM12a,RS14a}, this implies that each logical gate requires about 100 operations from the USFTO, for a total of $3\times 10^{12}$ for the entire algorithm. In turn, this implies that magic-state distillation must be accurate to at least 12 digits. Following convention and assuming that  noisy magic states can be prepared to accuracy 1\%, state-of-the-art distillation protocols \cite{MEK12a} require nearly $300$ such noisy input states to distill one of sufficient quality. Concluding this example, the total overhead associated to distillation and compilation is over $ 10^4$ magic-state inputs on average per logical gate (and even more Clifford operations, which are ignored in our analysis). This represents a major roadblock towards physical realization of fault-tolerant quantum computation.

To reduce this overhead, we employ a USFTO which is {\em over-complete}, in the sense that some operations could be removed from it without affecting its universality. However, removing such redundant gates would affect the synthesis cost. Specifically, our set comprises the Clifford gates (generated by controlled-not $\Lambda(X)$, Hadamard $H$, and phase gate $S$), and the infinite family $R_Y(\theta_k) = \exp(-i\pi Y/2^k)$, $k=3,4,\ldots$, which are rotations of angle $\theta_k = 2\pi/2^k$ around the $y$-axis of the Bloch sphere. Note that the cases $k=1$ and $k=2$ result in Clifford operations, while $k=3$ corresponds to the $T$ gate which is commonly used to complete the USFTO. Also, note that in concrete applications where each gate needs only be implemented to a desired accuracy $\delta$, we can effectively truncate the family since for large enough $k$, the rotation $R_Y(\theta_k)$ can be substituted by the identity to yield an error of magnitude $\delta \approx 2^{-k}$. For this reason we will limit our study to $k<90$ since larger values have no conceivable utility.

We realize the gates $R_Y(\theta_k)$ by distilling the associated magic states $|Y_k\rangle = \cos(\theta_k/2)|0\rangle + \sin(\theta_k/2)|1\rangle$. These states are injected into the quantum computation using the quantum circuit of Fig. 1 a), which consumes one state $|Y_k\rangle$ and Clifford operations to realize a rotation $R_Y$ by an angle $\pm \theta_k$. The sign of the rotation is completely random but known. This randomness doesn't really impact the synthesis cost as we now explain.

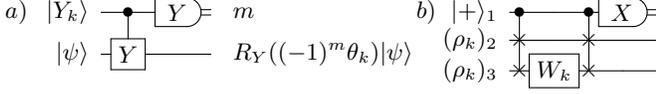
\begin{figure}

\centering
\[
a)\quad\quad\quad \Qcircuit @C=0.4em @R=.5em {
		\lstick{\ket{Y_k}}		& \ctrl{1}		& \measureD{Y} 	&  \cw	&\rstick{m}\\
		\lstick{\ket{\psi}}			& \gate{Y}		& \qw			& \qw	& \rstick{R_Y((-1)^m\theta_k)\ket{\psi}}
	}	
	\quad\quad\quad\quad\quad\quad\quad\quad b)\quad\quad\quad
	\Qcircuit @C=0.4em @R=.5em {
		\lstick{\ket{+}_1} 	&\ctrl{1}		& \qw			&\ctrl{1}		& \measureD{X}	& \cw \\
		\lstick{(\rho_k)_2}	& \qswap		& \qw			& \qswap		& \qw			& \qw \\
		\lstick{(\rho_k)_3}	& \qswap \qwx	&  \gate{{W}_k}		& \qswap \qwx	& \qw			& \qw	
	} 
 \]
	\caption{a) State injection circuit. The controlled qubit is initially prepared in magic state $|Y_k\rangle$ and the target is in an arbitrary state. Following the application of $\Lambda(Y)$, the measurement of the controlled qubit along the $y$ axis with outcome $m=\pm 1$ results in a rotation of $\pm \theta_k$ of the target qubit. b) Distillation circuit for $|Y_k\rangle$.  The $\Lambda({\rm SWAP})$ gates are used to project two noisy versions $\rho_k$ of $|Y_k\rangle$ onto the even-parity subspace, leading to a quadratic improvement of their accuracy.}
	\label{fig:Injection}
\end{figure}

Consider a single qubit rotation $U_{\hat n}(\theta)$ of angle $\theta$ around axis $\hat n$. Existing synthesis schemes can approximate this unitary transformation to absolute accuracy $\delta$ at cost $c\log^b(1/\delta)$ where $c$ and $b$ are some constants. As we now demonstrate, the compilation cost using our USFTO scales instead with the relative accuracy $\varepsilon = \delta/|\theta|$ as $3\log(6/\varepsilon)/2+3$. We decompose the gate using Euler angles as $U_{\hat n}(\theta) = R_Z(\alpha)R_Y(\beta)R_X(\gamma)=  HS^\dagger R_Y(\gamma)S H  R_Y(\beta) S^\dagger R_Y(\alpha)S$, so it requires six Clifford gates  and three $R_Y$ rotations of angles $|\alpha|, |\beta|, |\gamma| \leq 2|\theta|$ (see Supplementary Information),  each needing to be executed to relative accuracy $\varepsilon$/6. This means that each of these angles can be expressed with $\ell = \log(6/\varepsilon)$ significant bits. With $\alpha_k \in \{0,1\}$, $\alpha = \sum_{k=h}^{h+\ell} \alpha_k 2\pi/2^k$ is a rotation of magnitude $2^{-h}$ written to relative accuracy $2^{-\ell}$, and is straightforwardly expressed with at most $\ell$ gates from our USFTO as $R_Y(\alpha) =  R_Y(\theta_h)^{\alpha_h}R_Y(\theta_{h+1})^{\alpha_{h+1}}\ldots R_Y(\theta_{h+\ell})^{\alpha_{h+\ell}}$. These $\ell$ gates are executed sequentially, starting from $k=h+\ell$ down to $k=\ell$. At stage $k$ of this execution, suppose the state injection circuit Fig. 1 produces the outcome $-1$. The rotation should have been by angle $\theta_k$ but this outcome has produced $-\theta_k$ instead, so the state needs to be rotated by angle $2\theta_k = \theta_{k-1}$. This can be fixed by substituting $\alpha \leftarrow \alpha+ \theta_{k-1}$, and pursuing the execution of the circuit at $k-1$. Because of this intrinsic randomness, this execution will require $\ell/2+1$ gates on average.

We now explain how to distill the magic states $|Y_k\rangle$. Landahl and Cesare have proposed a distillation protocol for these states that uses a family of shortened Reed-Muller codes \cite{LC13a}. Unfortunately, the Reed-Muller codes are highly inefficient, so any synthesis overhead gained from this approach is overwhelmed by an increased distillation cost at high noise rate, although their approach can offer some advantages if the input magic states are of sufficiently high quality. Instead we build on a protocol introduced by Meier, Eastin, and Knill \cite{MEK12a} to distill $|Y_3\rangle$. Let us describe their protocol in the more general setting of current interest. The high-level distillation circuit for these states is shown at  Fig 1b), detailed implementtions are summarized in Fig. 2.

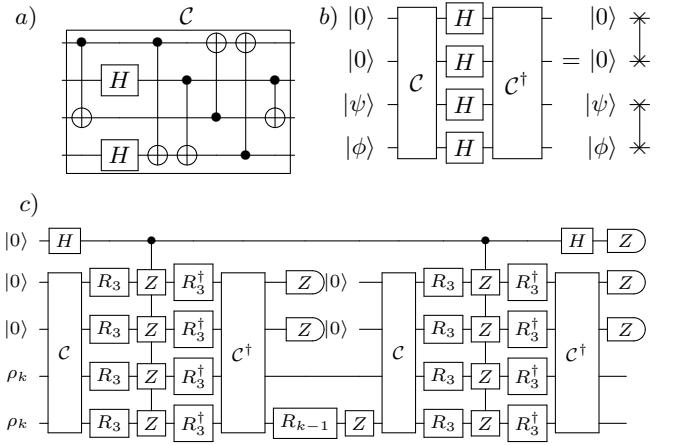
\begin{figure}
\[	a)\quad \Qcircuit @C=0.4em @R=.5em {
		& 		& 		& 		& \mbox{$\mathcal{C}$} & 	& 		& 		& \\
		& \ctrl{2}	& \qw	& \ctrl{3}	& \qw	& \targ	& \targ	& \qw	& \qw\\
		& \qw	& \gate{H}	& \qw	& \ctrl{2}	& \qw	& \qw	& \ctrl{1}	& \qw\\
		& \targ	& \qw	& \qw	& \qw	& \ctrl{-2}	& \qw	& \targ	& \qw\\
		& \qw	& \gate{H}	& \targ	& \targ	& \qw	& \ctrl{-3}	& \qw	& \qw \gategroup{2}{2}{5}{8}{1.25em}{-}
	}
\quad b) \quad\quad
	 \Qcircuit @C=0.4em @R=.5em {
		\lstick{\ket0}	& \multigate{3}{{\mathcal C}}	& \gate{H}	& \multigate{3}{{\mathcal C}^\dagger}	& \qw &  &     &&&&&&   \lstick{\ket0} 	&\qswap 		& \qw\\
		\lstick{\ket0}	& \ghost{{\mathcal C}}		& \gate{H}	& \ghost{{\mathcal C}^\dagger} 		& \qw	  &  & =  &&&&&&   \lstick{\ket0}	& \qswap \qwx 	& \qw\\
		\lstick{\ket{\psi}}& \ghost{{\mathcal C}}		& \gate{H}	& \ghost{{\mathcal C}^\dagger} 		& \qw 	  &  &     &&&&&&   \lstick{\ket\psi}	& \qswap 		& \qw\\
		\lstick{\ket{\phi}}& \ghost{{\mathcal C}}		& \gate{H}	& \ghost{{\mathcal C}^\dagger} 		& \qw 	  &  &     &&&&&&    \lstick{\ket\phi}	& \qswap \qwx 	& \qw\\
	} 
	\]
\vspace*{-0.5cm}
\begin{align*}
	c)& \hfill \\
&{\scriptsize \Qcircuit @C=0.4em @R=.5em {
		\lstick{\ket{0}}	& \gate{H}	 				&\qw				& \ctrl{1}		& \qw 				& \qw		 				& \qw				& \qw		& \qw					&\qw	& \ctrl{1}		&\qw			& \gate{H}				& \measureD{Z}	\\
		\lstick{\ket{0}}	& \multigate{3}{{\mathcal C}} 	& \gate{R_3}	& \gate{Z}		& \gate{R_3^\dagger}	& \multigate{3}{{\mathcal C}^\dagger}& \measureD{Z}		& \lstick{\ket{0}}	& \multigate{3}{{\mathcal C}}	& \gate{R_3}	& \gate{Z}		& \gate{R_3^\dagger}	& \multigate{3}{{\mathcal C}^\dagger}	& \measureD{Z}	\\
		\lstick{\ket{0}}	& \ghost{{\mathcal C}}		& \gate{R_3}	& \gate{Z}	\qwx	& \gate{R_3^\dagger}	& \ghost{{\mathcal C}^\dagger}		& \measureD{Z}		& \lstick{\ket{0}}	& \ghost{{\mathcal C}}		& \gate{R_3}	& \gate{Z} \qwx	& \gate{R_3^\dagger}	& \ghost{{\mathcal C}^\dagger}			& \measureD{Z}	\\
		\lstick{\rho_k}	& \ghost{{\mathcal C}}		& \gate{R_3}	& \gate{Z} \qwx	& \gate{R_3^\dagger} & \ghost{{\mathcal C}^\dagger}		& \qw				& \qw		& \ghost{{\mathcal C}}		& \gate{R_3}	& \gate{Z} \qwx	& \gate{R_3^\dagger}	& \ghost{{\mathcal C}^\dagger}			& \qw			\\
		\lstick{\rho_k}	& \ghost{{\mathcal C}}		& \gate{R_3}	& \gate{Z}	\qwx & \gate{R_3^\dagger}	& \ghost{{\mathcal C}^\dagger}		& \gate{R_{k-1}}	& \gate{Z}		& \ghost{{\mathcal C}}		& \gate{R_3}	& \gate{Z} \qwx	& \gate{R_3^\dagger}	& \ghost{{\mathcal C}^\dagger}			& \qw			
	}}
	\end{align*}
	\caption{ Detailed implementation of the distillation circuit. a) Encoding circuit for a 4-qubit error-detecting code. b) In this code, the Hadamard gate applied to every qubit has the effect of swapping the two encoded qubits (and the two ancillary qubits). c) Overall circuit combining the primitives of a), b) to implement the circuit of Fig. 1b). We denote $R_{k} = R_Y(\theta_k)$ for compactness.}
	\label{fig:distillationCircuit}
\end{figure}

Given the orthogonal basis $|Y_k\rangle$, $|\overline Y_k\rangle  = Y|Y_k\rangle =  i\sin(\theta_k/2)|0\rangle - i\cos(\theta_k/2)|1\rangle$, we define the phase flip operator $W_k = |Y_k\rangle\!\langle Y_k| -  |\overline Y_k\rangle\!\langle \overline Y_k|$. A direct calculation shows that $W_k = R_Y(\theta_{k-1})Z$, so the gate $W_k$ can be realized by injecting $|Y_{j}\rangle$ states with $j<k$. The circuit of Fig. 2.a) performs a measurement of the two-qubit `parity' operator $M_k = W_k\otimes W_k$.  To understand how this leads to error suppression, suppose for simplicity that each input qubit is prepared in the faulty magic state $\sqrt{1-\epsilon}|Y_k\rangle + \sqrt \epsilon_i |\overline Y_k\rangle$ (the argument generalizes to arbitrary forms of noise, see Supplementary Information), such that $\epsilon=0$ corresponds to a perfect magic state. Their joint state is
\begin{equation*}
(1-\epsilon)|Y_k, Y_k\rangle + \epsilon |\overline Y_k, \overline Y_k\rangle + \sqrt{\epsilon(1-\epsilon)} (| Y_k, \overline Y_k\rangle + |\overline Y_k,  Y_k\rangle).
\end{equation*}
The first two components of this state are $+1$ eigenstates of $W_k\otimes W_k$ since they have even parity, while the last two have eigenvalue $-1$ since they have odd parity. Thus, a measurement of $W_k\otimes W_k$ produces result $+1$ with probability $1-2\epsilon + 2\epsilon^2$, in which case the post-measurement state will be proportional to  $(1-\epsilon)|Y_k, Y_k\rangle + \epsilon |\overline Y_k, \overline Y_k\rangle$. Since the magnitude of the error component has decreased from $\sqrt \epsilon$  to $\epsilon$, we see that the error has been suppressed quadratically. On the other hand, the result $-1$ is obtained with complementary probability $2(\epsilon-\epsilon^2)$, in which case the qubits are discarded. 

An immediate difficulty with this distillation protocol is that the gate $\Lambda({\rm SWAP})$ it uses is not a Clifford transformation. To realize it, we encode a pair of qubits into a 4-qubit {\em error-detecting} code.  The Clifford circuit $\mathcal C$ of Fig. 2a) performs the encoding and maps the single-qubit Pauli operators $Z_i$ and $X_i$ as follows:
\begin{align}
(Z_1,X_1) &\rightarrow (ZZZZ, XZXZ)\label{eq:PauliMap}\\
(Z_2,X_2) &\rightarrow (XXXX, IZII)\\
(Z_3,X_3) &\rightarrow (ZIIZ,XIXI)\\
(Z_4,X_4) &\rightarrow (XIIX,ZIZI).
\end{align}
A key property seen in this transformation is that exchanging $X$'s for $Z$'s has the effect of swapping the last two lines of the equation, which corresponds to swapping the two encoded qubits. Since the Hadamard gate realizes the $X$-$Z$ exchange, we deduce that $H^{\otimes 4}$ performs the logical SWAP, cf. Fig. 2b).  

We can therefore substitute the $\Lambda({\rm SWAP})$ with four $\Lambda(H)$, but these are still not part of the Clifford group. However, using the identity $H = R_Y(\theta_3) Z R_Y(-\theta_3)$, we can express $\Lambda(H) = \Lambda[R_Y(\theta_3) Z R_Y(-\theta_3)] = R_Y(\theta_3) \Lambda(Z) R_Y(-\theta_3)$. We conclude that $\Lambda(H)$, and therefore $\Lambda({\rm SWAP})$, can be implemented with gates $R_Y(\theta_3)$ from our USFTO and $\Lambda(Z)$, which is a Clifford operation, cf. Fig. 2c). 

To recapitulate, the distillation of states $|Y_k\rangle$ requires 1) two noisy input states $|Y_k\rangle$, 2) one near-perfect collection of states $|Y_{j}\rangle$ for all $j<k$ used to implement the phase inversion gate $W_k = R_Y(\theta_{k-1})Z$, and 3) sixteen near-perfect states  $|Y_3\rangle$ used to implement the gates $\Lambda({\rm SWAP)}$. This distillation protocol lends itself to a recursive procedure \cite{GC99a} where previously distilled states $|Y_j\rangle$ for $j<k$ are used to distill states $|Y_k\rangle$.  

To get the recursion started at $k=3$, corresponding to the case studied in \cite{MEK12a}, requires additional analysis. Indeed, states $|Y_3\rangle$ are required to implement the $\Lambda({\rm SWAP})$, but only noisy $|Y_3\rangle$ states are available at this stage of the recursion. Note however that the gates $R_Y(\theta_3)$ are used inside an error-detecting code, cf. Fig. 2c), so they need not be perfect. A $-1$ measurement outcome at the output of the circuit $\mathcal C^\dagger$ reveals that one or more error has occurred in the execution of the encoded SWAP gate. Rejecting the instances where such a non-trivial error syndrome is found suppresses any first order error, thus preserving the quadratic error suppression of the ideal circuit of Fig. 1b). 

In general, we can express a noisy magic state $\rho_k$ in the $|Y_k\rangle,\ |\overline Y_k\rangle$ basis as
\begin{equation}
\rho_k = 
\left(\begin{array}{cc}
1-\epsilon_k & \Delta_k \\
\Delta_k^* & \epsilon_k
\end{array}\right)
\end{equation}
with $0\leq\epsilon_k\leq 1/2$ and $0\leq |\Delta_k|^2 \leq \epsilon_k-\epsilon_k^2$, the case $\Delta_k = \epsilon_k = 0$ corresponding to the perfect magic state $|Y_k\rangle$. For a fixed $k$ and given a set of input noise parameters $(\epsilon_3,\Delta_3, \epsilon_4, \Delta_k,\ldots, \epsilon_k, \Delta_k)$, using computer-assisted calculation we can derive an exact expression for 1) the average output noise parameter $(\epsilon_k', \Delta_k')$ of the distilled state, and 2) the expected number $N_j^k$ of consumed resource states $\rho_j$ of each kind $3\leq j\leq k$. The expectations are taken over the intrinsic randomness of the protocols, averaging over possible measurement outcomes. This calculation is realized by exactly simulating the quantum circuit of Fig. 2c), which is tractable since it only involves 5 qubits (see Supplementary Information). 

While these expressions are too lengthy to describe here, they will have the following form to leading order,
\begin{equation*}
\epsilon_k^{\rm out} \approx \epsilon_k^2 + 2\binom{8}{2}\epsilon_{3}^2 + \epsilon_{k-1} + \frac 12 \epsilon_{k-2} + \frac 14 \epsilon_{k-3}\ldots+\frac 1{2^{k-4}}\epsilon_3,
\end{equation*}
where it is implicitly assumed that $\epsilon_j = 0$ for $j<3$ (Clifford operations). The first term comes from the ideal distillation circuit of Fig. 1b) which quadratically reduces the error. The second term comes from the 8 $R_Y(\theta_3)$ gates used to implement the $\Lambda({\rm SWAP})$ inside the error-detecting code. It takes 2 faults out of these 8 gates to produce an undetected error. The extra factor of 2 accounts for the two occurrences of the $\Lambda({\rm SWAP})$ in the protocol. Finally, the last terms come from the $W_k = R_Y(\theta_{k-1})$ gate which consumes one $|Y_{k-1}\rangle$ state, consumes one $|Y_{k-2}\rangle$ state with probability $1/2$, etc. We note that in general, we can use $|Y_3\rangle$ states of different accuracies to implement the $W_k$ and the $\Lambda({\rm SWAP})$, since the latter appears inside a code, but we will omit this detail here for simplicity.

Similarly, we can estimate the expected number $N_j^k$ of states $\rho_j$ consumed during one distillation round of $|Y_k\rangle$ to be $N_j^k = [2^{j-k+1} + 16 \delta_{j,3}] \bar r$, where
$\bar r \approx  (1+  16 \epsilon_3 + 2\epsilon_k)$
is the average number of times the protocol needs to be repeated before all five measurement outcomes in Fig. 2c) return the value +1. A -1 outcome can be obtained either when one of the 16 $R_Y(\theta_3)$ gate is faulty or when one of the two input $|Y_k\rangle$ states are faulty. The behaviour and dependence of the off-diagonal terms $\Delta_k$ is more difficult to derive intuitively, but we note that their value had essentially no effect on the exact calculation; setting all $\Delta_j=0$ had no significant impact on our results.

To complete the analysis, we need a distillation schedule. To distill a state $|Y_k\rangle$ to accuracy $\delta$, our scheme makes use of previously distilled $|Y_j\rangle$ states  for $j< k$ with given noise parameters $(\epsilon_j, \Delta_j)$. To what accuracy should these resource states have been previously distilled? If they were not sufficiently distilled, their use in the distillation of $|Y_k\rangle$ could actually increase its error. On the other hand, using states $|Y_j\rangle$ that were distilled to a much greater precision than the targeted accuracy $\delta$ is wasteful. While we have not thoroughly optimized the distillation sequence, we used the following rule of thumb. A perfect distillation circuit  gives $\epsilon_k^{\rm out}=(\epsilon_k^{\rm in})^2$. With the use of imperfect magic states $|Y_j\rangle$, this output accuracy is instead $\epsilon_k^{\rm out} \approx (\epsilon_k^{\rm in})^2 + \sum_{j<k} \alpha_j \epsilon_j^{\beta_j}$ for some integers $\alpha_j$ and $\beta_j$.  Given this, we use resource states $|Y_j\rangle$ of accuracy $\epsilon_j \approx [(\epsilon_k^{\rm in})^2/\alpha_j]^{1/\beta_j}$. The intuition behind this rule is that each source of error will contribute equally to the output error $\epsilon_k^{\rm out}$. 

Figure 3 shows the overhead, defined as the number of noisy magic input states, required to distill a state $|Y_k\rangle$  to a desired accuracy $\delta$. Consistently with previous studies \cite{MEK12a,BH12b}, we assumed that the states $|Y_j\rangle$ can be prepared to accuracy 1\%. Note however that for $j>8$, the state  $|0\rangle$ is better than $1\%$ accurate approximation to $|Y_j\rangle$, and so we substituted all noisy input magic states by $|0\rangle$ for $j>8$. While we could have performed that substitution for all $j>3$ --- in which case $|Y_3\rangle$ states would have been the only non-Clifford inputs to our protocol ---  we obtained a slightly lower overhead with this prescription.

 \begin{figure}
\centering\includegraphics[scale=0.54]{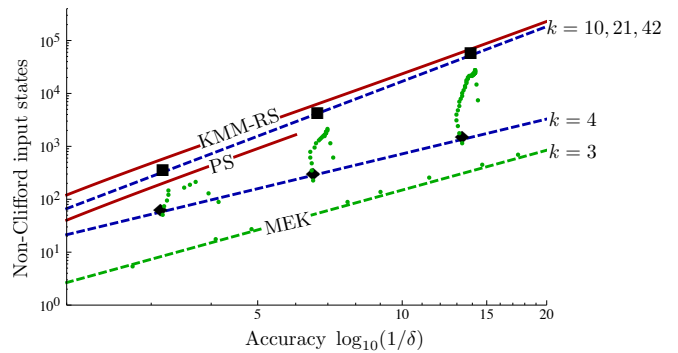}
\caption{Overhead,  measured as the number of input non-Clifford states, to realize a gate to accuracy $\delta$. The diamonds are the overhead to realize $R_Y(\theta_4)$.  Green dots  are distillation costs of $|Y_k\rangle$, with $4\leq k \leq \log_2(2\pi/\delta)$ (smaller angles $\theta_k < \delta$ can be substituted by 0 to accuracy $\delta$.)   Squares represent the cost of an arbitrary rotation of angle $\theta\leq 8\delta$. Red lines are obtained by combining the distillation scheme MEK of \cite{MEK12a} with the gate synthesis KMM-RS of either \cite{KMM12a} or  \cite{RS14a} (give very similar overhead), or PS of \cite{PS13a}. Note that of these three schemes, only \cite{RS14a} has an efficient compiling algorithm, so it directly compares to our approach. Due to its exponential cost, the compiler of \cite{PS13a} can only synthesize gates of accuracy $10^{-6}$ with reasonable computing power.}
\label{fig:Results}
\end{figure}

Returning to our opening example with a $\delta = 10^{-10}$ target accuracy, we see on Fig. 3 the $10^4$ overhead obtained by combining the distillation scheme \cite{MEK12a} and synthesis scheme \cite{KMM12a}. This overhead is to realize the target gate $R_Y(\pi/10)$, but we note that these protocols are largely insensitive to the target gate. In contrast, the overhead of our protocol depends on the target gate, and it ranges between 100-10,000 (green dots on Fig. 3) for the family of gates $R_Y(\theta_k)$, an improvement of up to 2 orders of magnitude (which moreover achieves a better accuracy  $\delta = 10^{-13}$). While this improvement is realized for specific single qubit gates $R_Y(\theta_k)$, we note that these gates occur very naturally in many quantum algorithms; for instance they are the only non-Clifford gates appearing in the quantum Fourier transform circuit \cite{NC00a}. For arbitrary  rotations, the overhead shown on Fig. 3  increases only with the logarithm of the relative accuracy of the rotation, as explained above. This also leads to substantial savings in many practically relevant settings, e.g., in quantum simulations \cite{Llo96b} where the Trotter-Suzuki formula is used to decompose the time-evolution operator into a product of small, accurate rotations, for which the overhead is minor. 

Lastly, for arbitrary rotations, our scheme does slightly worst than existing schemes. We note however that it can be generalized straightforwardly to the distillation of the family of state $|Y_k^m\rangle = |Y(m2\pi/2^k)\rangle$ for a fixed integer $0< m < 2^k$: distillation of $|Y_k^m\rangle$ can be realized given access to distilled $|Y_j^m\rangle$ with $j=3,4,\ldots k-1$. Since any rotation can be written as $m2\pi/2^k$ to $k$ bits of accuracy, this provides a way of realizing any rotation $R_Y(\theta)$ using on average only two magic states. This approach entirely avoids the need to compile (aside from an Euler angle decomposition), has a constant gate synthesis cost, and pushes all the gate synthesis overhead into the distillation. This last feature is important since distillation occurs off-line, i.e. it does not involve the data qubits.

In conclusion, we have presented a scheme to distill complex magic states which can offer significant savings over the traditional distillation/synthesis approaches. There are many foreseeable ways to obtain further gains from our scheme.  For instance, the rule of thumb we have used to determine the distillation schedule could be thoroughly optimized. The overhead of our scheme stems primarily from the implementation of the $\Lambda({\rm SWAP})$, which uses $16$ states $|Y_3\rangle$ to distill two qubits, a 2/16 yield. We have  investigated a generalization of our scheme based on a family of high rate codes that achieve a $m/4(m+1)$ yield for integer $m$, resulting in an additional 3-fold reduction of the overhead (see Supplementary Information). Further savings could be obtained using the approach of \cite{PS13a} to find more efficient distillation circuits. 

\medskip
\noindent{\em Acknowledgements---} This work was supported by Canada's NSERC and Qu\'ebec's FRQNT through the network INTRIQ. DP acknowledges the hospitality of The University of Sydney where this work was completed.

\bibliographystyle{/Users/dpoulin/archive/hsiam}
\bibliography{/Users/dpoulin/archive/qubib}

\newpage

\appendix

\section{Further improvements}

In this appendix we discuss possible ways of further reducing the compilation/distillation overhead of our scheme. From the start, we note that our protocol is equivalent to \cite{MEK12a} when distilling the $T$ gate (i.e. $k=3$), and that this forms the base of our recursion. The vast majority of studies on magic state distillation has focused on the $T$ gate, and any future improvement there can be directly incorporated into our protocol by substituting it for the first step of our recursion.  

One clear path to improvements is to use the same distillation tools with a thoroughly optimized distillation schedule. The rule of thumb we have used, which consists in setting the contribution to the final error from every noise source to be equal, ignores the fact that different components have different costs. Accounting for these costs would lead to a reduced overhead: the schedule should permit a larger contribution to the final error from a costly component. 

The central component of the distillation scheme is the controlled-SWAP gate. Following \cite{MEK12a}, it is implemented inside a 4-qubit quantum code where it can be substituted by 4 $\Lambda(H)$, and each of those can further be substituted by two $|Y_3\rangle$ state injections. Since one distillation round requires two $\Lambda({\rm SWAP})$ and distills two qubits, we obtain a rate of 1/8 distilled qubit per $|Y_k\rangle$ states consumed.  Increasing this rate would reduce the overhead.

For the distillation of states $|Y_k\rangle$ with $k>3$, a rate 1/7 can be obtained by realizing the $\Lambda({\rm SWAP})$ directly on the data, not making use of any code. Indeed, the  $\Lambda({\rm SWAP})$ can be realized with seven $T$ gates \cite{GKMR13a}. However, by doing so we would loose the benefit of the additional noise suppression offered by the code, so it is not obvious that this would be beneficial, at least early in the distillation schedule when the noise is relatively high.

It is possible to replace the 4-qubit code with a different code to achieve a rate $m/4(m+1)$, where $m$ is any positive integer. This is asymptotically a two-fold improvement of the yield, and given the recursive nature of our protocol this gain can be amplified to a more substantial value. Specifically, the code family has parameters $[[2m+2,2m, 2]]$; it is described by the stabilizer generators $Z^{2m+2}$ and $X^{2m+2}$ and has logical operators
\begin{align}
\overline Z_j &= \prod_{i=0}^{2j+1} Z_i, \ \ \overline X_j = X_{2j+1}X_{2j+2} \ \  {\rm for}\  0\leq j \leq m-1 \\ 
\overline Z_j &= \prod_{i=0}^{2j+1} X_i, \ \ \overline X_j = Z_{2j+1}Z_{2j+2} \ \ {\rm for}\  m\leq j\leq 2m-1.
\end{align}
The 4-qubit code used above corresponds to the special case $m=1$. Just like the 4-qubit code, this code has the property that swapping all $X$ and $Z$ operators, which is realized with the transversal Hadamard gate, has the effect of swapping pairs of logical qubits $j$ with $j+m$. Thus, this enables an $m$-fold parallelization of our original scheme at higher encoding  rate. Note however that this code still detects only a single error, and that by increasing $m$ we create  more opportunities for errors to accumulate. As a consequence, if each $|Y_3\rangle$ state used to implement the $\Lambda(H)$ is accurate to $\epsilon$, the probability of detecting an error scales like $m\epsilon$, and the probability of a harmful undetected error scales like $m^2\epsilon^2$. While this is a deterioration over the case $m=1$, it offers an additional flexibility in the distillation schedule that can be greatly beneficial, as we now explain.

As can be seen on the MEK data of Fig. 3, only very sparse values of the accuracy $\delta$ can be realized with standard distillation protocols. This is because the error is essentially squared at each iteration with a fixed pre-factor, i.e., $\epsilon \rightarrow c \epsilon^2$, leading to the discrete set of values $\epsilon, c\epsilon^2, c(c\epsilon^2)^2$, etc. This coarseness has the drawback that we will sometimes be forced to used un-necessarily accurate and costly gates, simply because there is a large gap in the range of available accuracies. This problem occurs not only during the implementation of the algorithm, but in the distillation procedure itself where previously distilled states $|Y_j\rangle$ are used to assist the distillation of $|Y_k\rangle$ with $k>j$. With the  enlarged code family proposed here, we can use the parameter $m$ to fine tune the accuracy of the distilled states. Combined with the improved rate, this has the potential to lead to substantial savings. Fig. 4 shows gains obtained by choosing the optimal value of $m$ at each step of the distillation, and demonstrates up to 3-fold improvement over the case $m=1$. 

We note that this calculation was realized using the leading order expansion described in the main text, adapted to the case $m>1$. Moreover, during the distillation of $|Y_k\rangle$, we assumed that states $|Y_j\rangle$ with $j<k$ of arbitrary accuracy $\epsilon_j$ could be accessed  at a cost $C_j(\epsilon)$, where this cost function was obtained by fitting a discrete set of achieved accuracies. In other words, we ignored the coarseness of the achievable accuracies. Since one of the main advantage of the schemes with $m>1$ is the possibility to finely tune the accuracy, we expect a more complete calculation to yield an even larger improvement over the $m=1$ case.

\begin{figure}
\centering\includegraphics[scale=0.6]{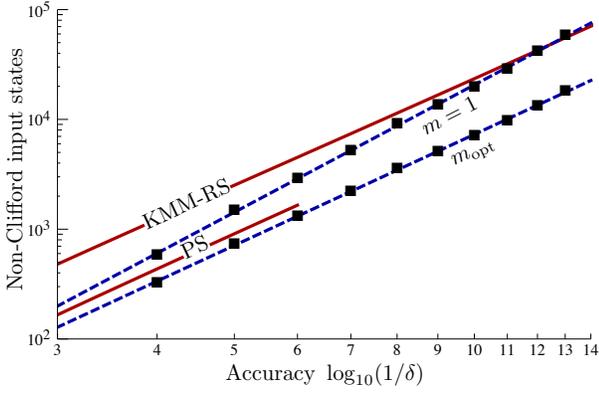}
\caption{As in Fig. 3. Approximate overhead of a rotation $R_Y(\theta)$ by angle $\theta < 8\delta$ (relative accuracy 1/8). Compares the original protocol $m=1$ with a protocol allowed to use higher rate codes $m_{\rm opt}$. For a precision of $\epsilon=10^{-13}$, a factor improvement of $\sim 3.2$ is observed. }
\label{fig:HigherYield}
\end{figure}

\section{Error Analysis}

In this appendix we give the details of the calculations and simulations for the circuits of Figure 2c).

\subsection{Imperfect $W_k$}
\label{app:imperfectWk}

Performing the phase flip operator $W_k$ requires the use of resource states $\ket{Y_{k'}}$ ($0\leq k'< k$) which are imperfect. First, the state $\ket{Y_{k-1}}$ is injected using the circuit of Fig.~1a). In order to account for errors in the magic state, we write down its effect (we label the top wire 1 and the bottom wire, 2)
\begin{align}
	\rho_{k-1}\otimes \rho&\rightarrow |\pm i\rangle\!\langle\pm i|_1\Lambda(Y_2)(\rho_{k-1}\otimes\rho) \Lambda(Y_2)|\pm i\rangle\!\langle\pm i|_1
\end{align}
where we drop normalization and where the $\pm$ sign is determined by the measurement outcome. If $\rho_{k-1}$ is perfect, then the circuit applies $R_Y(\pm\theta_{k-1})$ to $\rho_k$. Otherwise, recall that $\ket{\overline{Y}_{k-1}}=Y\ket{Y_{k-1}}$. Since, $Y=iXZ$ and that a) $Z_1$ commutes with any controlled unitary $\Lambda(U_2)$, and that b) $X_1$ propagates to $X_1\otimes U_2$ if $U_2$ is self-adjoint, then $Y_1 1\!\mathrm{l}_2$ propagates to $Y_1Y_2$ through $\Lambda(Y_2)$. Tracing over the resource state, the effect of injecting an imperfect state is

\begin{align}
	\rho & \xrightarrow{m=\pm1}	\\ &(1-\epsilon_{k-1})\rho_\pm+\epsilon_{k-1} Y\rho_\pm Y \pm \Delta_{k-1}Y\rho_\pm \pm \Delta_{k-1}^*\rho_\pm Y,\nonumber
\end{align}
where we take advantage of the fact that $[R_Y(\theta),Y]=0$ and where we have defined $\rho_+=R_Y(\theta_{k-1})\rho R_Y^\dagger(\theta_{k-1})$ corresponding to measurement outcome $+1$ and $\rho_-=R_Y(-\theta_{k-1})\rho R_Y^\dagger(-\theta_{k-1})$ corresponding to measurement outcome $-1$. The states $\rho_+$ and $\rho_-$ are obtained with equal probabilities. When $\rho_-$ is obtained, a $R_Y(\theta_{k-2})$ correction is required which involves another faulty resource state. Errors propagate again and combine with previous ones. However, errors are always of $Y$-type. For example, if we inject another rotation, in order to have no resulting error, the same error has to happen on both injections. Similarly a $Y\rho$ error in the first injection and a $\rho Y$ error in the second one results in a $Y\rho Y$ error, etc. We define the error amplitude vector $\vec\epsilon_k=(1-\epsilon_k,\epsilon_k,\Delta_k,\Delta_k^*)$ for imperfect resource state $\rho_k$. These combinations of errors define a product, noted $\times$, on such vectors:
\begin{align}
	& (a,b,c,d)\times(e,f,g,h)\overset{\textrm{def}}{=} \\	& (ae+bf+cg+dh, af+be+ch+dg, \nonumber\\
										& ag+bh+ce+df, ah+bg+cf+de). \nonumber
\end{align}
Using transpositions, $\tau_{ij}$, this can be rewritten
\begin{align}
	& A\times B \\&=(A\cdot B, A\cdot \tau_{12}\tau_{34}B, A\cdot \tau_{13}\tau_{24}B, A\cdot \tau_{14}\tau_{23}B),\nonumber
\end{align}
where ``$\cdot$" is the usual scalar product of vectors.

We note $\vec\epsilon_{k-}=(1-\epsilon_k,\epsilon_k,-\Delta,-\Delta^*)$, the error applied when the measurement outcome is $-1$. Using the vector product just defined, the imperfect application of $R_Y(\theta_{k-1})$ gives an expression for $\vec\epsilon_{W_k}$,
\begin{align}
	\vec \epsilon_{W_k}&=\frac 1 {2^{k-3}} (\prod_{j=3}^{k-1}\vec\epsilon_{j-})+\sum_{i=3}^{k-1} \frac 1 {2^{k-i}} (\prod_{j=i+1}^{k-1}\vec\epsilon_{j-}) \times \vec\epsilon_{i}.
\end{align}
The first term corresponds to having to apply all rotations $k>k'\geq 3$. This occurs with probability $1/2^{k-3}$. The second term sums over all other possibilities: for a given value $i$, we assume injections $k>k'>i$ have measurement outcome $-1$ and that the $i$th measurement outcome is +1. This happens with probability $1/2^{k-i}$.

Using the resulting error vector $\vec\epsilon_{W_k}$, the imperfect phase operator as the following effect (omitting normalization)
\begin{align}
	\tilde{W_k}(\rho) \rightarrow	&(\vec \epsilon_{W_k})_1W_k\rho W_k+(\vec \epsilon_{W_k})_2 YW_k\rho W_k Y\nonumber\\
							& - (\vec \epsilon_{W_k})_3YW_k\rho W_k - (\vec \epsilon_{W_k})_4W_k\rho W_k Y.ª
\end{align}

\subsection{$\Lambda$(Swap) gadget}

The circuit of Figure 2.c) shows that a $\Lambda({\rm Swap})$ gadget uses eight $R_{Y}(\theta_3)$ rotations, each requiring a $\ket{Y_3}$ injection. As explained in appendix B.1, errors in the resource state translate into errors on the target qubit. However, since $W_3=H$ (Clifford), we can perfectly twirl all  $\ket{Y_3}$ states such that the error is always diagonal: $\rho_3=(1-\gamma)|Y_3\rangle\!\langle Y_3|+\gamma|\overline{Y}_3\rangle\!\langle \overline{Y}_3|$. Consequently, we account for errors in the $\Lambda({\rm Swap})$ gadget by following each $R_{Y_3}$ rotations by a $Y$-Pauli channel of strength $\gamma$: $\rho\rightarrow(1-\gamma)\rho+\gamma Y\rho Y$.

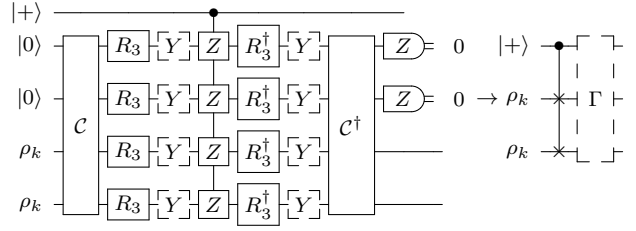
\begin{figure}
{\footnotesize
	\[\Qcircuit @C=0.4em @R=.5em {
		\lstick{\ket{+}}	& \qw					&\qw				& \qw			& \ctrl{1}		& \qw 				& \qw			& \qw							&\qw \\
		\lstick{\ket{0}}	& \multigate{3}{{\mathcal C}} 	& \gate{R_{3}}	& \gateBox{Y}{--}	& \gate{Z}		& \gate{R_{3}^\dagger}	& \gateBox{Y}{--}	& \multigate{3}{{\mathcal C}^\dagger}	& \measureD{Z}	& \cw	& \rstick{0}	& & & & &				& & & & & & \lstick{\ket{+}}	& \qw	& \ctrl{1}		& \qw	& \multigateBox{2}{\Gamma}{--}	& \qw\\
		\lstick{\ket{0}}	& \ghost{{\mathcal C}}		& \gate{R_{3}}	& \gateBox{Y}{--}	& \gate{Z}	\qwx	& \gate{R_{3}^\dagger}	& \gateBox{Y}{--}	& \ghost{{\mathcal C}^\dagger}			& \measureD{Z}	& \cw	&\rstick{0}		& & & & & \rightarrow	& & & & & & \lstick{\rho_k}	& \qw	& \qswap		& \qw	& \ghost{\Gamma}			& \qw\\
		\lstick{\rho_k}	& \ghost{{\mathcal C}}		& \gate{R_{3}}	& \gateBox{Y}{--}	& \gate{Z} \qwx	& \gate{R_{3}^\dagger} & \gateBox{Y}{--}	& \ghost{{\mathcal C}^\dagger}			&\qw				&\qw		&\qw			& & & & &				& & & & & & \lstick{\rho_k}	& \qw	& \qswap\qwx	& \qw	& \ghost{\Gamma}			& \qw\\
		\lstick{\rho_k}	& \ghost{{\mathcal C}}		& \gate{R_{3}}	& \gateBox{Y}{--}	& \gate{Z}	\qwx & \gate{R_{3}^\dagger}	& \gateBox{Y}{--}	& \ghost{{\mathcal C}^\dagger}			&\qw				& \qw	&\qw
	}\]
}
	\caption{Passing the noise through the gadget. The dashed boxes represent the resulting effective stochastic noise maps.}
\end{figure}

To keep the simulation on only three qubits, we push these errors through the gadget and project on the trivial syndrome, as is shown in Fig.~5. In the following, we label wires 0 to 4 from top to bottom. To propagate $Y_i$ errors through a $\Lambda_0(H_i)$, we note that $\{Y_i,\Lambda_0(H_i)\}=0$ implies $Y_i\Lambda_0(H_i)=\Lambda_0(H_i)Z_0Y_i$. Then, we can propagate $Y_i$ errors through the decoding circuit $\mathcal{C}^\dagger$, using \eq{PauliMap}. By determining the effect of every single-qubit $Y$ error this way, we can deduce the effective stochastic map  $\Gamma$, see Fig.~5. Note that the way it is now defined, $\Gamma$ is not trace preserving. We define $Pr^{(1,2)}_{\Lambda(\mathrm{Swap})}=\mathrm{Tr}(\Gamma(\rho^{(1,2)}))$, the probability of measuring the trivial syndrome in the first or second $\Lambda(\mathrm{Swap})$ gadget.


\subsection{Error Calculation and Simulation Details}

Using the results of the two previous subsections, we calculate the single-qubit output state as a function of the various inputs in the following manner:

\begin{enumerate}
	\item Set $\rho=|+\rangle\!\langle+|\otimes\rho_k\otimes\rho_k$
	\item Set $\rho^{(1)}=\Gamma(\Lambda(\mathrm{Swap})\rho\Lambda(\mathrm{Swap})^\dagger)$
	\item Compute $\mathrm{Pr}^{(1)}=\mathrm{Tr}(\rho^{(1)})$
	\item Normalize $\rho^{(1)}\leftarrow\rho^{(1)}/\mathrm{Pr}^{(1)}$
	\item Set $\rho^{(W_k)}=\tilde W_k (\rho^{(1)})$
	\item Set $\rho^{(2)}=\Gamma(\Lambda(\mathrm{Swap})\rho^{(W_k)}\Lambda(\mathrm{Swap})^\dagger)$
	\item Compute $\mathrm{Pr}^{(2)}=\mathrm{Tr}(\rho^{(2)})$
	\item Normalize $\rho^{(2)}\leftarrow\rho^{(2)}/\mathrm{Pr}^{(2)}$
	\item Post-select on measuring the trivial syndrome: $\rho^{\rm out}=\mathrm{Tr}_{1,2}(|+\rangle\!\langle+|\otimes1\!\mathrm l\otimes 1\!\mathrm l \rho^{(2)})$
	\item Compute $\mathrm{Pr}^{\rm out}=Tr(\rho^{\rm out})$
\end{enumerate}

From $\rho^{\rm out}$, we can extract $\epsilon^{\rm out}=\epsilon^{\rm out}(\gamma,\epsilon_k,\delta_k)$ and $\Delta^{\rm out}=\Delta^{\rm out}(\gamma,\epsilon_k,\delta_k)$. Using these quantities, we define the cost of a distillation round:

\begin{align}
	\mathrm{Cost}_k(\epsilon^{\rm out}&,\Delta^{\rm out})= \\
	&\frac 1 2 \bigg(\frac{2 \mathrm{Cost}_k(\epsilon_k)+ 8 \mathrm{Cost}_3(\epsilon_{3,\Lambda{\rm (Swap)}})}{\mathrm{Pr}^{(1)}}\nonumber\\
	& + \mathrm{Cost}_{W_k}(\{\epsilon_{k'},\Delta_{k'}\})\nonumber\\
	&+8 \mathrm{Cost}_3(\epsilon_{3,\Lambda{\rm (Swap)}})\bigg) \frac{1}{\mathrm{Pr}^{(2)}\mathrm{Pr}^{\rm out}}\nonumber
\end{align}
where $3\leq k'<k$. The $1/2$ prefactor accounts for the fact that we distill two copies of $\rho_k$. The cost to apply a $W_k$ operator is the sum of the cost of all states $3\leq k'<k$ multiplied by the probability of being needed:
\begin{align}
	\mathrm{Cost}_{W_k}(\{\epsilon_{k'},\Delta_{k'}\})=	& \sum_{i=3}^{k-1}\frac{\mathrm{Cost}_i(\epsilon_i,\Delta_i)}{2^{k-1-i}}.
\end{align}

For $3\leq k\leq 8$, we initialize the cost to $\mathrm{Cost}_k(0.01,0)=1$ and the state to $\rho_k=0.99|Y_k\rangle\!\langle Y_k|+ 0.01|\overline{Y}_k\rangle\!\langle \overline{Y}_k|$. For $k\geq9$, the state $\ket0$ is actually closer in trace distance to the resource state than an imperfectly prepared version with $1\%$ error. In this case, we initialize $\rho_k=|0\rangle\!\langle0|$ and $\mathrm{Cost}_k(\epsilon_{k,0},\Delta_{k,0})=0$, where $\epsilon_{k,0}$ and $\Delta_{k,0}$ are obtained by expressing $|0\rangle\!\langle0|$ in the basis defined by $\ket{Y_k}$. We find $\epsilon_{k,0}=\sin^2(\theta_k/2)$ and $\Delta_{k,0}=\sin(\theta_k)/2$. In this case, the error is purely off-diagonal, i.e. it saturates the bound $|\Delta|^2\leq\epsilon-\epsilon^2$ derived from the positivity of $\rho_k$.

For $k=3$, the cost table is given by the scheme of \cite{MEK12a}. Starting at $k=4$, two copies $\mathrm{Cost}_4(0.01,0)=1$ are used to distill two improved copies. This gives a new value $\mathrm{Cost}_k(\epsilon_{k,1},\Delta_{k,1})=\mathrm{Cost}_k(\epsilon^{\rm out},\Delta^{\rm out})$. Using two copies of this improved state, another round of distillation is performed and so on. Remember that the precisions and costs of the inputs are chosen according to the rule of thumbs discussed earlier (see main text). Once a set of values has been obtained for the distillation of $\ket{Y_k}$, $\ket{Y_{k+1}}$ can be distilled and so on, bootstrapping on the protocol of \cite{MEK12a}.


\section{Euler angle decomposition}

In this section we demonstrate the claim made in the main text that a small-angle single-qubit rotation $R_{\hat n}(\theta)$ can be decomposed into a sequence of rotations $R_Z(\alpha)R_Y(\beta)R_X(\gamma)$ around the three axis of the Bloch sphere, all with angles of magnitude bounded by $2\theta$. On one hand, in the axis-angle representation, the rotation matrix takes the form
\begin{widetext}
\begin{equation}
R_{\hat n}(\theta) = \left(
\begin{array}{ccc}
\cos \theta +n_x^2 \left(1-\cos \theta\right) & n_x n_y \left(1-\cos \theta\right) - n_z \sin \theta & n_x n_z \left(1-\cos \theta\right) + n_y \sin \theta \\ 
n_y n_x \left(1-\cos \theta\right) + n_z \sin \theta & \cos \theta + n_y^2\left(1-\cos \theta\right) & n_y n_z \left(1-\cos \theta\right) - n_x \sin \theta \\ 
n_z n_x \left(1-\cos \theta\right) - n_y \sin \theta & n_z n_y \left(1-\cos \theta\right) + n_x \sin \theta & \cos \theta + n_z^2\left(1-\cos \theta\right) 
\end{array}
\right).
\end{equation}
On the other hand, in the Euler angle decomposition, this rotation matrix is
\begin{equation}
R_Z(\alpha)R_Y(\beta)R_X(\gamma) = \left(
\begin{array}{ccc}
\cos\beta \cos\alpha & \cos\gamma \sin\alpha + \sin\gamma \sin\beta \cos\alpha &   \sin\gamma \sin\alpha - \cos\gamma \sin\beta \cos\alpha \\
  -\cos\beta \sin\alpha &  \cos\gamma \cos\alpha - \sin\gamma \sin\beta \sin\alpha & \sin\gamma \cos\alpha + \cos\gamma \sin\beta \sin\alpha \\
  \sin\beta             &  -\sin\gamma \cos\beta                                          &   \cos\gamma \cos\beta 
\end{array}
\right).
\end{equation}
\end{widetext}
By changing orientation of $\hat n$, we can assume without loss of generality that $0\leq \theta\leq \pi/8$. Equating these two matrices and noting that $|n_x|,\ |n_y|,\ |n_z| \leq 1$, element (3,1) yields the inequality
\begin{align}
|\sin\beta| \leq 1-\cos\theta + \sin\theta \leq \sin 2\theta,
\end{align}
which implies that $|\beta| \leq 2\theta$. We proceed similarly for the other angles. Equating element (2,1) of these matrices yields 
\begin{align}
|\sin\alpha| &\leq (1-\cos\theta + \sin\theta)/|\cos\beta| \\
 &\leq (1-\cos\theta + \sin\theta)/\cos2\theta  \leq \sin 2\theta,
\end{align}
which implies that $|\alpha| \leq 2\theta$. Finally, the bound on $\gamma$ is obtained following an identical reasoning using element $(3,2)$ of the matrix equality. 


\end{document}